\documentstyle[12pt]{article}

\begin{document}

\title{STANDARD PHYSICS SOLUTION TO THE SOLAR NEUTRINO PROBLEM ?}
\author{Arnon Dar\\
Department of Physics \\
and \\	Asher Space Research Institute \\
Technion-Israel Institute of Technology \\
 Haifa 32000, Israel}
\maketitle 

\begin{abstract} 

The $^8$B solar neutrino flux predicted by the standard solar model (SSM)
is consistent within the theoretical and experimental uncertainties with
that observed at Kamiokande. The Gallium and Chlorine solar neutrino
experiments, however, seem to imply that the $^7$Be solar neutrino flux is
strongly suppressed compared with that predicted by the SSM. If the $^7$Be
solar neutrino flux is suppressed, still it can be due to
astrophysical effects not included in the simplistic SSM.  Such effects 
include short term fluctuations or periodic variation of the 
temperature in the solar core, rotational mixing of $^3$He in the solar
core, and dense plasma effects which may strongly enhance p-capture by
$^7$Be relative to e-capture. The new generation of solar observations
which already look non stop deep into the sun, like Superkamiokande
through neutrinos, and SOHO and GONG through acoustic waves, may point at
the correct solution. Only Superkamiokande and/or future solar neutrino
experiments, such as SNO, BOREXINO and HELLAZ, will be able to find out
whether the solar neutrino problem is caused by neutrino properties
beyond the minimal standard electroweak model or whether it is just a 
problem of the too simplistic standard solar model.

\end{abstract}

\section{Introduction} 
\bigskip 

Solar neutrinos have been detected on Earth in four pioneering solar
neutrino ($\nu_\odot$) experiments, in roughly the expected numbers,
demonstrating that the sun is indeed powered by fusion of hydrogen into
helium. However the precise counting rates in the $\nu_\odot$ experiments
(e.g., Hampel, this proceedings)
are approximately one half that predicted by the Standard Solar Model
(SSM). This discrepancy, which has persisted for 30 years, has become
known as the solar neutrino problem (SNP). It has attracted much attention
of both astrophysicists and particle physicists for two main reasons. 
First, astrophysicists were surprised to find out that the sun, and in
particular the nuclear reactions in its core, are not accurately described
by the simple SSM. Second, particle physicists found that natural 
extensions of the minimal 
standard electroweak model (SEM) can solve elegantly the SNP. However when
astronomers had a closer look at the sun through helioseismology, X-ray
and UV observations it turned out to be ``a bewildering turmoil of
complex phenomena'', showing unexpected features and behavior at any
scale. It has a strange complex internal rotation, unexplained magnetic
activity with unexplained 11 year cycle, unexpected anomalies in its
surface elemental abundances, unexplained explosions in its atmosphere and
unexplained mechanism that heats its million degree corona and accelerates
the solar wind.  Perhaps the surface of the sun is complex because we
can see it and the center of the sun is not only because we cannot?
Perhaps the SSM which has been improved continuously over the past three
decades (see e.g., Clayton 1968, Bahcall 1989), which still assumes an
exact spherical symmetry, no mass loss or mass accretion, no angular
momentum loss or gain, no differential rotation and zero magnetic field
through the entire solar evolution, is a too simplistic picture and does
not provide a sufficiently accurate description of the core of the sun 
and/or the neutrino producing reactions there?

Indeed, here I will show that the solar neutrino problem does
not provide solid evidence for neutrino properties beyond the SEM 
and that standard physics solutions to the SNP are possible. In 
particular I will argue that:

\noindent
1. There is no $^8$B  solar neutrino problem.

\noindent
2. Only observations of spectral distortions and/or flavor change of
 $\nu_\odot$'s in future $\nu_\odot$ experiments, like Superkamiokande,
SNO, Borexino and HELLAZ may establish that neutrino properties beyond the
SEM are responsible for the SNP. 

\noindent
3. There is no solid evidence for a real deficit of $^7$Be solar neutrinos

\noindent
4. A real deficit of $^7$Be solar neutrinos, if there is one, 
may still be explained by standard physics and/or astrophysics.

\noindent
5. The major three new solar observatories, which are already running and
looking into the solar interior, the Superkamiokande solar neutrino
observatory that began taking data on April 1, 1996, the solar
heliospheric observatory (SOHO) that was launched on December 2, 1995 and
is now observing the sun non stop, and the ground based telescopes in the
Global Oscillations Network Group (GONG) which have just begun observing
solar oscillations around the clock, may very soon point at the correct
solution to the SNP (for general reviews see Science, 31 May 1996). 

\section{Is There a $^8$B Solar Neutrino Problem?}
\smallskip

Table I presents a comparison between the solar neutrino observations and
the SSM predictions of Bahcall and Pinsonneault 1995 (BP95) and of Dar and
Shaviv 1996 (DS96). Although BP (and some other similar SSM calculations) 
predict a $^8$B solar neutrino flux that is approximately 2.4 larger than
observed by Kamiokande (see Hampel, this proceedings), DS 
predict a flux
consistent with that observed by Kamiokande. The differences between BP
and DS are summarized in Table II (for details see Dar and Shaviv 1996). 
The difference between the predicted $^8$B flux are mainly due to the use
of updated nuclear reaction rates by DS, differences in the calculated
effects of diffusion, differences in the initial solar composition assumed
in the two calculations and the use of an improved equation of state by
DS. They reduce the predicted $^8$B flux by approximate factors of 0.55,
0.81, 0.95 and 0.96, respectively (the remaining differences are mainly
due to inclusion of partial ionization effects, premain sequence evolution
and deviations from complete nuclear equilibrium by DS 
which were neglected by BP, and due to different numerical
methods, fine zoning and time steps used in the two calculations):

\noindent
{\bf Nuclear Reaction Rates:} The uncertainties in the nuclear reaction
rates at solar conditions are still large due to (1) uncertainties in the
measured cross sections at laboratory energies, (2) uncertainties in their
extrapolations to solar energies, (3) uncertainties in dense plasma
effects (screening, correlations and fluctuations) on reaction rates. 
Rather than averaging measured cross sections that differ by many standard
deviations, DS used for the extrapolation only the most recent and
consistent measurements of the relevant nuclear cross sections. Because
sub-Coulomb reactions take place when the colliding nuclei are far apart,
the Optical Model and the Distorted Wave Born Approximation give a
reliable description of their energy dependence.  DS have used them for
extrapolating measured sub-Coulomb cross sections to solar energies.  BP
preferred to rely on published extrapolations of averaged cross sections
based on energy dependences calculated from sophisticated microscopic
nuclear reaction models (e.g. Johnson et al 1992).
Similar screening corrections (which by
accidental cancellation have a very small net effect on
$\phi_{\nu_\odot}(^8{\rm B})$) have been used by DS and BP. The updated
``astrophysical S factors'' which were used by DS are listed in Table
II. They reduce the BP predictions by approximately a factor of 0.55 . 

\noindent
{\bf Diffusion:} Diffusion, caused by density, temperature, pressure,
chemical composition and gravitational potential gradients play an
important role in the sun since it modifies the local chemical composition
in the sun. The relative changes in SSM predictions due to diffusion of
all elements are summarized in Table III. While BP found a rather large
increases in the predicted $^7$Be, $^8$B, $^{13}$N, $^{15}$O and $^{17}$F
solar neutrino fluxes; 14\%, 36\%, 52\%, 58\%, and 61\% which result in
36\%, 33\%, 9\% increases in their predicted rates in Kamiokande, Homestake,
and in GALLEX and SAGE, respectively, DS found only a moderate increase
due to diffusion, 4\%, 10\%, 23\%, 24\% and 25\%, respectively, in the
above fluxes, which result in 10\%, 10\% and 2\% increase in the predicted
rates in Kamiokande, Homestake, and in GALLEX and SAGE, respectively.
Although the two diffusion calculations assumed a different initial solar
chemical composition (see below) and BP approximated the diffusion of all
elements heavier than $^4$He by that of fully ionized iron (the DS
calculations followed the diffusion of each element separately and used
diffusion coefficients calculated for the actual ionization state of each
element at each shell in the sun as obtained from solving the local Saha
equations), these cannot fully explain the above large differences. 
Recent independent diffusion calculations by Richard et al. (1996) 
obtained similar results to those obtained by DS as can be
seen from Table III (we interpolated the results from 
the two models of Richard et al. to the initial chemical 
composition assumed by DS).   

\noindent 
{\bf Initial Chemical Composition:} The initial chemical composition
influences significantly the solar evolution and the present density,
chemical composition and temperature in the solar core, which determine
the solar neutrino fluxes.  In particular, the calculated radiative
opacities, which in turn determine the temperature gradient in the solar
interior, are very sensitive to the heavy elements abundances (the heavy
elements are not completely ionized in the sun).  Apart from the noble
gases, only a few elements such as H, C, N and O, which were
able to form highly volatile molecules or compounds,  have escaped complete
condensation in primitive early solar system meteorites 
(see, e.g., Sturenburg
and Holweger 1990).  Thus, the initial solar abundances of all other
elements are expected to be approximately equal to those found in type I
carbonaceous chondrites as a result of their complete condensation in the
early solar system.  Since the chemical composition of the solar surface
is believed to have changed only slightly during the solar evolution (by
nuclear reactions during the Hayashi phase, by diffusion and turbulent
mixing in the convective layer during the main sequence evolution, and by
cosmic ray interactions at the solar surface) it has been expected that
the photospheric abundances of these elements are approximately equal to
those found in CI chondrites.  Over the past decades there have been many
initial disagreements between the meteoritic and photospheric abundances.
In nearly all cases, when the atomic data were steadily improved and the
more precise measurements were made, the photospheric values approached
the meteoritic values. The photospheric abundances are now as a rule in
very good agreement with the meteoritic values (Grevesse and Noels 1991;
1993). 
Since the  meteoritic values represent the initial values and are known
with much better accuracy (often better than 10\%) than the photospheric
ones, DS assumed that the initial solar heavy metal abundances are given
approximately by the meteoritic (CI chondrites) values of Grevesse and
Noels (1993) and adjusted the initial CNO and Ne abundances to reproduce
their observed photospheric abundances.  Also the unknown initial $^4$He
solar abundance has been treated as an adjustable parameter.  DS
``predicted'' its present photospheric mass fraction to be $ Y=0.238\pm
0.05$ in good agreement with the $^4$He surface mass fraction inferred
from helioseismology:  $Y_s=0.242\pm 0.003$ (Hernandez and
Christensen-Dalsgaard 1994. However, their formal error is highly
misleading because of the great sensitivity of the result to the model of
the solar atmosphere, the equation of state there and the atmospheric
opacities. We estimate that at present the $^4$He mass fraction at the
solar surface is not known from helioseismology better than $Y_s=0.242\pm
0.010$). BP adjusted the initial solar composition to reproduce 
the present day surface abundances which, except for the CNO and the nobel
gases, were assumed to be represented by their meteoritic values.

The photospheric abundances of $^7$Li, $^9$Be and $^{11}$B are smaller by
a factor of nearly 150, 3 and 10, respectively, than their meteoritic
abundances. The origin of such large differences is still not clear.  They
 cannot be explained by nuclear burning during the Hayashi phase although
significant Lithium burning does takes place during this phase. They may
be explained by rotational mixing (e.g., Richard et al 1996). Although the
initial solar (meteoritic) abundances of Lithium, Beryllium and Boron are
very small and do not play any significant role in solar evolution their
depletion perhaps can provide a clue to the real history of the convection
zone and the sun. 

\noindent
{\bf Equation of State:} The equation of state is used to calculate
the local density and temperature required to balance the gravitational
pressure in the sun. Since the neutrino producing reactions in the sun   
depend strongly on temperature, their predicted fluxes depend
strongly on the equation of state. DS have used an updated equation
of state which is described in detail in DS96. It is consistent with the
new OPAL equation of state (Rogers et al. 1996)
The use of an improved equation
of state reduce significantly our 1994 solar neutrino fluxes and 
improves the agreement between the sound speed 
in the solar core that we calculated from our SSM and the 
sound speed that is extracted from helioseismology. The
agreement with the updated sound speed from helioseismology (Christensen
Dalsgaard, 1996) is better than $2\times 10^{-3}$, 
as is 
demonstrated in Fig. 1. It is significantly better than the agreement 
obtain/reported by other SSM calculations.

\section{Evidence For $\nu_e$ Properties Beyond the Minimal SEM?}
\bigskip

Counting rates in $\nu_\odot$ experiments are formally given by
\begin{equation}
R=N_A\Sigma_i \phi_{\nu_\odot}(i)\int_{E_0}(dn_{\nu_i}/dE)\sigma_{\nu A}
(E)dE 
\end{equation}
where $N_A$ is the number of ``active'' atoms in the detector,
$\sigma_{\nu A}(E)$ is their cross section for neutrinos with energy E,
$dn_{\nu_i}/dE$ is the normalized energy spectrum of neutrinos
from reaction $i$ in the sun and $\phi_{\nu\odot}$ is their total flux.
Both, $dn_{\nu_i}/dE$ and $\sigma_{\nu A}$ follow directly from
the standard electroweak theory and are independent of the sun. 
($dn_{\nu_i}/dE$ is practically the standard $\beta$-decay
spectrum for the $\beta$-decays 2p$\rightarrow De^+\nu_e$,
$^8$B$\rightarrow 2\alpha e^+\nu_e$, $^{13}$N$\rightarrow
^{13}$C$e^+\nu_e$ and $^{15}$O$\rightarrow^{15}$N$e^+\nu_e$ and is a 
$\delta$-function for the electron captures 
$e^7$Be$\rightarrow\nu_e^7$Li
and $pep\rightarrow D\nu_e$.) Thus {\it conclusive evidence}
for new electroweak physics can be provided only by detecting
at least one of the following signals:

1. Spectral distortion of the fundamental $\beta$-decay spectrum.

2. Solar neutrino flavors other than $\nu_e$.         

3. A clear violation of the luminosity sum rule.

4. Rates which require negative $\phi_{\nu_\odot(i)}$. 

\noindent
So far, no such  clear evidence has been provided by the $\nu_\odot$
experiments. 

\noindent
{\bf Spectral Distortions:} Until recently, only Kamiokande could test
whether the spectrum of their detected $\nu_\odot$'s is consistent with
the $\nu_e$ spectrum from $\beta$-decay of $^8$B. Kamiokande observed an
electron recoil spectrum from $\nu_\odot e$ interactions which is
consistent, within their limited statistics, with that expected from an
undistorted $^8$B solar neutrino spectrum. Superkamiokande, which has been
running since April 1, 1996, will soon have  much larger statistics
(See Hampel these proceedings). 

\noindent {\bf Neutrino Oscillations}: Neutrino oscillations or neutrino
helicity flip can explain the solar neutrino observations (see Petcov,
this proceedings). However, no time variation which is predicted by a
magnetic helicity flip has been detected by the $\nu_\odot$ experiments,
nor could the experiments detect (Homestake, GALLEX and SAGE) or
distinguish (Kamiokande) between different neutrino flavors.
Superkamiokande will soon examine with a high level of 
sensitivity (real time, high
statistics) whether the $^8$B solar neutrino flux is time dependent while
only future experiments like SNO will be able to detect other neutrino
flavors (the sensitivity of Superkamiokande to temporal variation  
in the solar neutrino flux will be 
demonstrated by measuring the annual variation of the flux
due to the annual variation of the distance of
Earth from the sun).  

\noindent 
{\bf The Solar Luminosity Sum Rule:} If the sun derives its energy from
fusion of Hydrogen into Helium and {\it if it is in a steady state} where
its nuclear energy production rate equals its luminosity, then
conservation of baryon number, electric charge, lepton flavor and energy
requires that the total solar neutrino flux at Earth satisfies (e.g., Dar
and Nussinov 1991):
\begin{equation} 
\phi_{\nu_\odot}={2L_\odot\over
Q-2\bar{E}_\nu}~{1\over 4\pi D^{2}} \geq 6.52\times 10^{10}~cm^{-2}
s^{-1}~, 
\end{equation} 
where $D\approx 1.496 \times 10^{13}~cm$ is the distance to the sun,
$Q=26.733~MeV$ is the energy released when four protons fuse into Helium,
$\bar{E}_\nu=\sum{E_{\nu_i}}\phi_{\nu_i }/ \sum{\phi_{\nu_i}}$ is
the average energy of solar neutrinos and $\bar{E}_\nu\geq 0.265~MeV$
if the pp reaction in the sun produces $\nu_\odot$'s with the smallest
average energy.
Eq. (2) can be rewritten as a luminosity sum rule: 
\begin{equation} 
\Sigma_i(Q/2-\bar{E}_{\nu_i})\phi_{\nu_i}= S, 
\end{equation} 
where $S=L_\odot/ 4\pi D^2=1367~W~m^{-2}$ is the solar ``constant''. 
A clear Violation of eq. (2) or the solar luminosity sum rule,
can prove that lepton flavor is not conserved. In this conference the
Gallium experiments with the low energy threshold of 233 keV, which makes
them sensitive to almost all the SSM neutrinos, reported updated
time-averaged capture rates of $70\pm 8~SNU $ in GALLEX (Hampel, this
proceedings) and $72\pm 12~ SNU$ in SAGE (see Hampel, this proceedings). 
These new smaller rates  
are still consistent within the experimental uncertainties with $76\pm
2~SNU$, the ``minimal'' signal expected from eq. (2) and $\sigma_{Ga}=
(1.18\pm 0.02)\times 10^{45}~cm^{-2}$, if all the $\nu_\odot$'s were pp
$\nu$'s.  However, the $^8$B solar neutrino flux measured in Kamiokande,
$\phi_{\nu_\odot}=(2.8\pm 0.4)\times 10^6~cm^{-2}$, contributes another
$7\pm 2~SNU$ which increase the minimal expected signal in 
Gallium to
$83\pm 3~$ SNU.  This somewhat larger rate is still consistent within
$2\sigma$ with the capture rates measured by GALLEX and SAGE, in
particular if their rates are ``recalibrated'' by their new Cr source
experiments (see Hampel, this proceedings). But the Gallium 
experiments leave no room for significant
(SSM-like) contributions from $^7$Be and CNO solar neutrinos. This
confirms the combined results from the Chlorine experiment at Homestake
(see Hampel, this proceedings) and the Kamiokande experiment (see 
Hampel, this proceedings): 

\noindent
{\bf The Missing $\nu_\odot$'s:} Althogh the $^{37}$Cl experiment with an
energy threshold of $814~keV$ is completely blind to the pp solar
neutrinos it is sensitive to both the $^8$B neutrinos and the lower energy
$pep$, CNO and $^7$Be neutrinos. However, while the expected signal from a
$^8$B solar neutrino flux alone as measured by Kamiokande is $3.08\pm
0.53~SNU$, the time-averaged counting rate in the $^{37}$Cl
experiment is $2.56\pm 0.25~SNU$ (see Hampel, these proceedings. Although the
$^{37}$Cl experiment has not been ``calibrated'' with a neutrino source,
the Cr source experiments of GALLEX and SAGE suggest that the accuracy of
the radiochemical experiments is probably of the order of 10\%, or better.
Consequently, although the joint results from Homestake and Kamiokande do
not provide solid evidence for ``new electroweak physics'' (e.g.,
Bahcall and Bethe 1991) they do indicate that the combined
contributions from $^7$Be, CNO and pep solar neutrinos is strongly
suppressed in $^{37}$Cl compared with their SSM estimated contribution.

\smallskip

\section{Are $^7$Be Solar Neutrinos Missing?} 
\bigskip
Electron capture by $^7$Be into the ground state of $^7$Li produces 862
keV neutrinos. The threshold energy for neutrino absorption by $^{37}$Cl
is 814 keV. Thus, absorption of $^7$Be neutrinos by $^{37}$Cl produces 48
keV electrons.  The maximum energy of the pp solar neutrinos is 420 keV.
The threshold energy for neutrino absorption in $^{71}$Ga $(3/2^-)$ is 233
keV into the ground state $(1/2^-)$ and 408 into its first excited state
$(5/2^-)$. The produced electrons have therefore energies below 187
and 12 $keV$, respectively. If the theoretical cross sections for neutrino
absorption near threshold overestimate significantly their true values
then the predicted rates will significantly overestimate the expected
signals in the Chlorine and Gallium experiments.

An indication that final state interactions effects are not completely
understood is provided by Tritium $\beta$-decay.  Although final state
interactions in Tritium $\beta$-decay have been studied extensively,
they do not explain well the end-point $\beta$-decay spectrum ($E_e\sim
18.6~keV$). In all recent measurements, the measured spectrum yields a
negative value for the fitted squared mass of the electron neutrino
(see, e.g., Holzschun, this proceedings). 
Final state interactions effects (screening of the nuclear charge by
atomic electrons, exchange effects, radiative corrections, nuclear recoil
against the electronic cloud, etc) in neutrino captures near threshold in
$^{37}$Cl and $^{71}$Ga may be much larger because their Z values are much
larger and because the de Broglie wave lengths of the produced electrons
are comparable to the Bohr radii of the atomic K shells in Cl and Ga. If
final state interactions reduce considerably the near threshold absorption
cross sections of pp neutrinos in $^{71}$Ga (making room for the expected
contribution of $^7$Be solar neutrinos in Gallium) and of $^7$Be
neutrinos in $^{37}$Cl, perhaps they can make the solar neutrino
observations of Kamiokande and the Homestake experiment compatible.
Such an explanation of the solar neutrino problem implies that experiments
such as BOREXINO and HELLAZ will observe the full $^7$Be solar neutrino
flux.

\section{Astrophysical Solutions To The SNP}
\bigskip
 
Even if the $^7$Be solar neutrino flux is strongly suppressed, it does not
eliminate standard physics solutions to the solar neutrino problem: 

The ratio between the fluxes of $^7$Be and $^8$B solar neutrinos 
is given by  
\begin{equation} 
R={\phi_{\nu_\odot}(^7{\rm Be})\over \phi_{\nu_\odot}(^8{\rm B})}
 = {\int n_e n_7<\sigma v>_{e7}4\pi r^2dr\over
   \int n_p n_7<\sigma v>_{p7}4\pi r^2dr}.
\end{equation}
Because of the decreasing temperature and Be7 abundance as function of
distance from the center of the sun on the one hand, and the $\sim r^2$ 
increase in radial mass on
the other, the production of $^7$Be and $^8$B solar neutrinos
in the SSM peaks around an effective radius, $r_{eff}\approx
0.064R_\odot$ ($r_{eff}$ is approximately the radius within which 50\% of
the flux is produced) . The SSM also predicts a ratio of electron to
proton densities near the center of the sun, $n_e/n_p\sim 2$, consistent
with helioseismology observations.  Consequently, the SSMs predict  
\begin{equation}
R\approx {2<\sigma v>_{e7}\over <\sigma v>_{p7}} \approx 4.4\times 
10^{-17}S_{17}T_7^{1/6}e^{47.625/T_7^{1/3}},
\end{equation}      
where  $T_7$ is the temperature in $10^7K$ at 
the effective radius and   
$S_{17}$ is in $eV~barn$ units.
The SSMs yield $T_7(r_{eff})\approx 1.45$. Using $S_{17}(0)=17~eV~b$  
and  $\phi_\odot(^8{\rm B})=2.8\times 10^6~cm^{-2}~s^{-1}$ as observed
by Kamiokande, one can reproduce the SSM prediction (e.g., Dar and 
Shaviv 1996) 
\begin{equation}
\phi_{\nu_\odot}(^7{\rm Be})=R 
\phi_{\nu_\odot}(^8{\rm B})\approx 3.7\times 10^9~cm^{-2}~s^{-1}. 
\end{equation}

Astrophysical solutions of the solar neutrino problem aim towards
suppressing the value of R. Three alternatives are currently investigated:
 
\noindent 
{\bf Plasma Physics Effects}: The effects of the surrounding plasma on
nuclear reaction rates in dense stellar plasmas, and in particular on
proton and electron capture by $^7$Be in the sun are known only
approximately. In order to explain the deficit of $^7$Be solar neutrinos,
without much affecting the SSM, plasma screening effects must
reduce/enhance considerably electron/proton capture by $^7$Be, respectively,
relative to the predictions of the weak screening theory (Salpeter and Van
Horne 1969). This possibility is currently studied, e.g., by Shaviv and
Shaviv (1996) using numerical methods and by Brown and Sawyer 
(1996) using quantum
statistical mechanics techniques. Because of accidental cancellations the
screening corrections to the rates of all nuclear reactions do not change
the predicted $^8$B solar neutrino flux, but perhaps a more exact
treatment of screening may change R considerably. 
 
In principle, collective plasma physics effects, such as very strong
magnetic or electric fields near the center of the sun, may polarize the
plasma electrons, and affect the branching ratios of electron capture by
$^7$Be (spin $3/2^-$) into the ground state (spin $3/2^-$,
$E_{\nu_e}=0.863~MeV$, BR=90\%) and the excited state (spin $1/2^-$,
$E_{\nu_e}=0.381~MeV$, BR=10\%) of $^7$Li. Since solar neutrinos with
$E_{\nu_e}=0.381~MeV$ are below the threshold (0.81 MeV) for capture in
$^{37}$Cl and have a capture cross section in $^{71}$Ga that is smaller by
about a factor of 6 relative to solar neutrinos with
$E_{\nu_e}=0.863~MeV$, therefore a large suppression in the branching
ratio to the ground state can produce large suppressions of the $^7$Be
solar neutrino signals in $^{37}$Cl and in $^{71}$Ga. However, such an
explanation require anomalously large fields near the center of the sun. 
 
\noindent
{\bf Temporal and Spatial Variations in T:}
Davis (1996) has been claiming persistently that the solar neutrino
flux measured by him and his collaborators in the $^{37}$Cl radiochemical 
experiment is varying with time. Because of the possibility that
neutrinos may have anomalous magnetic moments, much larger than those
predicted by minimal extensions of the standard electroweak model, 
which can solve the solar neutrino problem (see Ould-Saada, these
proceedings), attention has been focused on anticorrelation between
the solar magnetic activity (the 11 year cycle) and the $\nu_\odot$
flux (see, e.g., Davis 1996). Also a day-night effect (e.g., Cribier et al
1986; Dar and Mann 1987)
due to resonant conversion of the lepton flavor of solar neutrinos
which cross Earth at night before reaching the solar neutrino detector
was not found by Kamiokande. However, the basic general question
whether the solar neutrino flux varies on a short time scale, has not been
fully answered, mainly because of the limited statistics of the
first generation of solar neutrino experiments.

The SSM predict no significant variation of the solar neutrino flux
on time scales shorter than millions of years. However, the sun has 
a differential rotation. It rotates once in $\sim$ 25 days near the 
equator, and in $\sim$ 33 days near the poles. Moreover, the observed
surface rotation rates of young solar-type stars are up to 50 times 
that of the sun. It suggest that the sun has been loosing angular
momentum over its lifetime. The overall spin-down of a sun-like
star by mass loss and electromagnetic radiation is difficult to
estimate from stellar evolution theory, because it depends on
delicate balance between circulations and instabilities that 
tend to mix the interior and magnetic fields that retard or
modify such processes. It is quite possible that the differential 
rotation extends deep into the core of the sun and causes there
spatial and temporal variations in the solar properties due to  
circulation, turbulences and mixing. Since R is very sensitive 
to the temperature, even small variations in temperature
can affect R significantly without affecting significantly   
the pp solar neutrino flux (the $^7$Be and $^8$B solar neutrinos
will come mainly from temperature peaks, while the pp neutrinos
will reflect more the average temperature).

In fact, a cross correlation analysis of the various data sets from
the Homestake, Kamiokande, GALLEX and SAGE, shows
an unexpected correlation: If arbitrary time lags are added
to the different solar neutrino 
experiments, the cross correlation is maximal
when these time lags vanish.  Moreover, a power spectrum analysis
of the signals shows a peak around 21 days, suggesting a periodical 
variation (Sturrock and Walther 1996). The effect may 
be a statistical fluke. However, it can also indicate
a real short time scale variation in the solar core. Fortunately, 
Superkamiokande will soon provide the answer to whether the $^8$B solar
neutrino flux is time-dependent or not. Relevant information
may come soon also from SOHO and GONG.   

\noindent
{\bf Mixing of $^3$He:} 
The SSM $^3$He equilibrium abundance increases sharply with
radius. Cumming and  Haxton  (1996) have recently suggested that
the $^7$Be
solar neutrino problem  could be circumvented in models
where $^3$He is transported into the core 
in a mixing pattern involving rapid filamental flow downward.
We note that if this mixing produces 
hot spots (due to enhanced energy release) they can increase 
the effective temperature for p capture by $^7$Be in a cooleenvironment,
reducing R while keeping the $^8$B solar neutrino flux at the 
observed level. Perhaps, helioseismology will be able to test that.  

Cummings and Haxton (1996) also noted that such mixing will have other 
astrophysical 
consequences. For example, galactic evolution models predict
$^3$He abundances in the presolar nebula and in
the present interstellar medium (ISM) that are substantially (i.e., a
factor of five or more) in excess of the observationally
inferred values.  This enrichment of the ISM is driven by low-mass
stars in the red giant phase, when the convective envelope reaches
a sufficient depth to mix the $^3$He peak, established during
the main sequence, over the outer portions of the star.  The
$^3$He is then carried into the ISM by the red giant wind.
The core mixing  
lowers the main sequence $^3$He abundance at large r. 
 
\section {The MSW Solution}
Standard solar models, like the one presented in this work,
perhaps can explain the results reported by
Kamiokande. However, standard physics cannot explain an $^{37}$Ar
production rate in $^{37}$Cl smaller than that expected from
the solar $^8$B neutrino flux measured by Kamiokande. If the
experimental results of Kamiokande and Homestake are interpreted
as an evidence for such a situation (e.g., Bahcall 1994; 1995),
they do imply  new physics beyond the standard particle physics
model (Bahcall and Bethe 1991). In that case
an elegant solution to the solar neutrino anomaly is
resonant neutrino flavor conversion in the sun, first proposed
by Mikheyev and Smirnov (1986) (see also Wolfenstein 1978; 1979).
It requires only a natural extension of the minimal standard
electroweak theory.  It is based on a simple quantum mechanical effect.
Many authors have carried out extensive calculations to determine
the neutrino mixing parameters
which can bridge between the predictions of the
standard solar models and the solar neutrino observations. They
found that  a neutrino mass difference
$\Delta m^2\sim 0.7\times 10^{-5}~eV^2 $ and a neutrino mixing of
$sin^2 2\theta\approx 0.5\times 10^{-2}$ can solve the solar
neutrino problem (see Petcov, these proceedings). These parameters, however,
cannot explain the neutrino-oscillation-like signal which was reported
by the LSND experiment (see Caldwell, these proceedings).

\section{Conclusions}
The solar neutrino problem may be an astrophysical problem.
An indication for that may come from observation of unexpected 
temporal variability of the solar neutrino flux by Superkamiokande
or from helioseismology observations by SOHO and GONG. 
An indication may also come from
cross correlation analysis of the time dependent of the counting rates
in GALLEX and Sage 
and of the counting rates of Kamiokande and Homestake. Such cross 
correlation analysis may test 
whether the time variation of the counting rates is statistical
or physical. Deviations of the experimental results from those predicted by
the standard solar models may reflect the approximate
nature of the models (which neglect angular momentum
effects, differential rotation, magnetic
field, angular momentum loss and mass loss during evolution
and do
not explain yet, e.g., solar activity and the surface depletion of
Lithium, Berilium and Boron relative to their meteoritic values,
that may or may not be relevant to the solar neutrino problem).
Improvements of the standard solar model should continue.
In particular,
dense plasma effects on nuclear reaction rates and radiative
opacities,
which are not well understood, may  affect the SSM
predictions
and should be further studied, both theoretically and experimentally.
Relevant information may be obtained from studies of thermonuclear
plasmas in inertial confinement experiments. Useful information
may also be obtained from improved data on screening effects
in low energy nuclear cross sections of ions, atomic beams and
molecular
beams incident on a variety of gas, solid and plasma targets.
 
Better knowledge of low energy nuclear cross sections is
badly needed. Measurement of crucial low energy nuclear cross
sections by new methods, such as measurements of the
cross sections for the radiative captures ${\rm p+^7Be\rightarrow
^8B+\gamma}$ and ${\rm ^3He+^4He\rightarrow ^7Be+\gamma}$ by
photodissociation of $^8$B  and $^7$Be in the coulomb field of
heavy nuclei are badly needed in order to determine whether there
is a $^8$B solar neutrino problem. 
 
The $^{37}$Ar production rate in $^{37}$Cl indeed may be smaller
than that expected from the flux of standard solar neutrinos as
measured
by electron scattering in the Kamiokande experiment. In that case
neutrino oscillations, and in particular the MSW effect, may
be the correct solution to the solar neutrino problem. Only future
experiments, such as SNO, Superkamiokande, BOREXINO and HELLAZ,
will be able to supply
a definite proof that Nature has made use of this beautiful effect.

\noindent
{\bf Acknowledgement:} This talk is based on an ongoing collaboration 
with Giora Shaviv.
\vfill
\eject
\bigskip
\section{ REFERENCES}
\smallskip
 
 \noindent
Anders, E. and Grevesse, N., 1989, Geochim. Cosmochim. Acta ,
{\bf 53}, 197. 
 
 \noindent
Bahcall, J.N. 1989, {\it Neutrino Astrophysics} (Cambridge
Univ. Press 1989).
 
 \noindent
Bahcall, J.N. \& Bethe, H. 1991, Phys. Rev. D. {\bf 44}, 2962.
 
 \noindent
Bahcall, J.N. \& Pinsonneault, M. 1995, Rev. Mod. Phys., (submitted).

  \noindent
Caldwell, D. 1996, these proceedings
 
  \noindent
Brown, L.S. \& Sawyer, R.F. 1996, preprint astro-ph 9610256

  \noindent
Clayton, D. 1968, {\it Princ. of Stellar Evolution \& Nucleosyn.}
(McGraw-Hill)

  \noindent
Christensen-Dalsgaard, J. 1996, Nucl. Phys. B (Proc. Suppl.) {\bf 48}, 325

  \noindent
Christensen-Dalsgaard, J. 1996a, private communication to G. Shaviv 

  \noindent
Cummings, A. \& Haxton, W. 1996, preprint nucl-th 9608045
 
  \noindent
Cribier, M. et al. 1986, Phys. Lett. {\bf B182 2}, 89

  \noindent
Dar, A. and \& Mann, A. 1987, Nature {\bf 325}, 790
 
  \noindent
Dar, A. and Nussinov, S. 1991, Particle World {\bf 2}, 117

  \noindent
Dar, A. \& Shaviv, G., 1996, ApJ, {\bf 468}, 933  
 
  \noindent
Dar, A. \& Shaviv G. 
1994, Proc. VI Int. Conf. on Neutrino Telescopes
(ed.. M. Baldo-Ceolin) p. 303. 

  \noindent
Davis, R. Jr. 1996, Nucl. Phys. B (Proc. Suppl.) {\bf 48}, 284

  \noindent
Dzitko, H. et al., 1995, ApJ. {\bf 447}, 428 (1995)
 
  \noindent
Grevesse, N., 1991, A\&A, {\bf 242}, 488
 
  \noindent
Grevesse, N. \& Noels, A., 1993, in {\it Origin and Evolution of the
Elements} (eds. N. Prantzos et al., Cambridge Univ. Press) p. 15
 
  \noindent
Grevesse, N. \& Noels, A., 1993, Phys. Scripta {\bf T47} , 133
 
  \noindent
Hampel, W. 1996, these proceedings and references therein.

  \noindent
Hata, N. et al., 1994, Phys. Rev. D {\bf 49}, 3622
 
 \noindent 
Hata, N. \& Langacker, P., (1995), Phys. Rev. D {\bf 52}, 420
 
  \noindent
Hernandez, E.P. \& Christensen-Dalsgaard, J., (1994), MNRAS
{\bf 269}, 475
 
  \noindent
Holzschun, E. 1966, these proceedings and references therein. 

  \noindent
Johnson, C.W. et al. 1992, ApJ. {\bf 392}, 320
 
  \noindent
Kovetz, A. \& Shaviv, G., 1994, ApJ. {\bf 426}, 787
 
  \noindent
Mikheyev, P. \& Smirnov, A. Yu. 1985, Yad. Fiz. {\bf 42}, 1441

  \noindent
Petcov, S. 1996, these proceedings and references therein.

  \noindent
Richard, O. et al. 1996, Submitted to A\&A, preprint astro-ph 9601136 

  \noindent
Rogers, F.J. et al. 1996, ApJ, {\bf 456}, 902

  \noindent
Salpeter, E.E. \& Van Horne, H.M. 1969, ApJ, {\bf 155}, 183  

  \noindent
Shaviv, G. \& Shaviv, N., 1996, ApJ, {\bf 468},
 
  \noindent
Sturenburg, S. \& Holweger, H., 1990, A\&A {\bf 237}, 125

  \noindent
Sturrock, P.A. and Walther, G. 1996, report astro-ph 9609152 
 
  \noindent
Wolfenstein, L. 1978, Phys. Rev. {\bf D17}, 2369
 
  \noindent

Wolfenstein, L. 1979, Phys. Rev. {\bf D20}, 2634

\noindent
{\bf Figure Caption}: Comparison between the sound speed in
the core of the sun as extracted from recent helioseismology
studies (Christensen Dlsgaard 1996a) and the sound speed calculated
from the standard solar model of Dar and Shaviv (1996).  

\clearpage 
 
{\bf Table Ia:} Comparison between the solar neutrino fluxes predicted
by the SSM of BP95 and of  
DS96, and measured by the four solar neutrino experiments.
$$\matrix{~~~~~\nu~{\rm Flux}\hfill & {\rm BP95} \hfill & {\rm   DS96}\hfill 
&{\rm  Observations} \hfill &{\rm Experiment} \hfill \cr
{\rm \phi }_{\nu }(pp)~[{10}^{10}{cm}^{-2}{s}^{-1}] \hfill &5.91
\hfill &6.10 \hfill  &\mit &\rm \cr
{\phi }_{\nu }(pep)~[{10}^{8}{cm}^{-2}{s}^{-1}] \hfill &
1.39\hfill &1.43\hfill &\rm \hfill &\rm \cr
{\phi }_{\nu }(^{7}{Be})~[{10}^{9}{cm}^{-2}{s}^{-1}]\hfill &5.18\hfill
&3.71\hfill &\rm \hfill
&\rm \hfill \cr
{\phi }_{\nu }(^{8}{B})~[{10}^{6}{cm}^{-2}{s}^{-1}]\hfill &6.48\hfill
& 2.49 \hfill &\rm 2.80\pm 0.40\hfill &\rm Kamiokande\cr
{\phi }_{\nu }(^{13}{N})~[{10}^{8}{cm}^{-2}{s}^{-1}]\hfill &6.4\hfill
& 3.82 \hfill &\rm \hfill &\rm \cr
{\phi }_{\nu }(^{15}{O})~[{10}^{8}{cm}^{-2}{s}^{-1}]\hfill &5.15\hfill
& 3.74 \hfill &\rm \hfill &\rm \cr
{\phi }_{\nu }(^{17}{F})~[{10}^{6}{cm}^{-2}{s}^{-1}]\hfill &6.48\hfill
& 4.53 \hfill &\rm \hfill &\rm \cr
\hfill& \hfill & \hfill &\hfill\cr
\Sigma (\phi \sigma)_{Cl}~[SNU]\hfill&  9.3\pm 1.4\hfill &
4.1\pm 1.2\hfill & 2.56\pm 0.25
 \hfill &{\rm Homestake}\hfill \cr
\Sigma(\phi\sigma)_{Ga}~[SNU]\hfill & 137\pm 8\hfill
& 115\pm 6\hfill
&70\pm 8 \hfill &{\rm GALLEX}\cr
\Sigma(\phi\sigma)_{Ga}~[SNU]\hfill & 137\pm 8\hfill
& 115\pm 6\hfill
&72\pm 12 \hfill &{\rm SAGE}\cr}$$

{\bf Table Ib} Characteristics of the BP95, DS94, and DS96
Solar Models in Table Ia
(c=center; s=surface; bc=base of convective zone;
${\rm \bar N=log([N]/[H])+12)}$.
 
$$\matrix{{\rm Parameter}\hfill& {\rm BP95}\hfill& {\rm
DS94}\hfill& {\rm
DS96}\hfill \cr
{T}_{c}~[{10}^{7}K] \hfill &1.584 \hfill &1.554 \hfill &1.561 \hfill \cr
{\rho }_{c}~[g~c{m}^{-3}]\hfill&156.2 \hfill&155.3\hfill&155.4\hfill
\cr
{X}_{c}\hfill&0.3333\hfill &0.3462\hfill &0.3424 \hfill \cr
{Y}_{c}\hfill&0.6456 \hfill &0.6359 \hfill &0.6380 \hfill \cr
{Z}_{c}\hfill&0.0211\hfill&0.01950 \hfill
&0.01940 \hfill \cr
{R}_{conv}~[R/R_{\odot}]\hfill&0.712 \hfill &0.7105 \hfill &0.7130
\hfill \cr
{T}_{bc}~[{10}^{6}{\rm K}]\hfill&2.20 \hfill &2.10 \hfill &2.105
\hfill \cr
{X}_{s}\hfill&0.7351 \hfill &0.7243 \hfill &0.7512 \hfill \cr
{Y}_{s}\hfill&0.2470 \hfill &0.2597 \hfill &0.2308 \hfill \cr
{Z}_{s}\hfill&0.01798 \hfill &0.01574 \hfill
&0.0170 \hfill \cr
\overline{N}_s{(^{12}C})\hfill&8.55\hfill &8.50 \hfill&8.55 \hfill \cr
\overline{N}_s{(^{14}N})\hfill&7.97 \hfill &7.92 \hfill&7.97\hfill \cr
\overline{N}_s{(^{16}O})\hfill&8.87 \hfill &8.82 \hfill&8.87 \hfill \cr
\overline{N}_s{(^{20}Ne})\hfill&8.08 \hfill &8.03 \hfill&8.08 \hfill
\cr
{T}_{eff}~[{\rm K}]\hfill& \hfill &5920 \hfill &5803
\hfill \cr}$$
 
\clearpage 
 
{\bf Table II:} Comparison between the SSM of Bahcall and Pinsonneult
(1995) and of Dar and Shaviv (1996).
$$\matrix{  
\hfill& {\rm ~~~~~~BP95}\hfill& {\rm 
~~~~~~DS96}\hfill \cr
\hfill&\hfill& \hfill \cr
{M}_{\odot} \hfill &1.9899\times 10^{33}~g \hfill &1.9899\times 10^{33}~g 
\hfill\cr
{L}_{\odot} \hfill&3.844\times 10^{33}~erg~s^{-1} \hfill& 3.844\times 10^{33}
~erg~s^{-1}\hfill\cr
{R}_{\odot}\hfill&6.9599\times 10^{10}~cm \hfill &6.9599\times 
10^{10}~cm\hfill \cr 
{t}_{\odot}\hfill&4.566\times 10^9~ y \hfill &4.57\times 10^9~y 
\hfill \cr 
{\rm Rotation}\hfill&{\rm Not~Included} \hfill & {\rm Not~Included} 
\hfill \cr 
{\rm Magnetic~Field}\hfill&{\rm Not~Included} \hfill & {\rm Not~Included} 
\hfill \cr 
{\rm Mass ~Loss}\hfill&{\rm Not~Included} \hfill & {\rm Not~Included} 
\hfill \cr 
{\rm Angular~Momentum~Loss}\hfill&{\rm Not~Included} \hfill & {\rm Not 
~Included} \hfill \cr 
{\rm Premain~Sequence~Evolution}\hfill&{\rm Not~ 
Included} \hfill & {\rm Included} \hfill \cr 
{\rm Initial~Abundances: }\hfill& \hfill & \hfill \cr
{\rm ^4He}\hfill& {\rm Adjusted~Parameter} \hfill &{\rm 
Adjusted~Parameter} \hfill \cr 
{\rm C,N,O,Ne}\hfill& {\rm Adjusted~Photospheric} \hfill &{\rm 
Adjusted~Photospheric} \hfill \cr 
{\rm All~Other~Elements}\hfill& {\rm Adjusted~``Photospheric''} \hfill &{\rm 
Meteoritic} \hfill \cr 
{\rm Photospheric~Abundances: }\hfill&\hfill & \hfill \cr
{\rm ^4He }\hfill&{\rm Predicted} \hfill&{\rm Predicted}\hfill \cr
{\rm C,N,O,Ne}\hfill& {\rm Observed} \hfill &{\rm Observed} \hfill \cr 
{\rm All~Other~Elements}\hfill& {\rm =~Meteoritic} \hfill &{\rm 
Predicted} \hfill \cr 
\hfill& \hfill & \hfill \cr
{\rm Radiative~Opacities}\hfill&{\rm OPAL~1994}
\hfill & {\rm OPAL~1996} \hfill \cr 
{\rm Equation~ of~ State}\hfill&{\rm Straniero ~1988?}
\hfill & {\rm DS~ 1996} \hfill \cr 
{\rm Partial~ Ionization~ Effects}\hfill&{\rm Not~ 
Included} \hfill & {\rm Included} \hfill \cr 
{\rm Diffusion~ of~ Elements:}\hfill& \hfill & \hfill \cr
{\rm H,~^4He}\hfill&{\rm Included} \hfill&{\rm Included} \hfill \cr
{\rm Heavier~Elements}\hfill&{\rm Approximated~ by~ Fe } \hfill&{\rm 
 All~Included} \hfill \cr
{\rm Partial~ Ionization~ Effects}\hfill&{\rm Not~Included } \hfill&{\rm 
 Included} \hfill \cr
{\rm Nuclear~ Reaction~ Rates:}\hfill& \hfill & \hfill \cr 
S_{11}(0)\hfill &
\rm 3.896\times {10}^{-22}~keV\cdot b \hfill &\rm 4.07\times {10}^{-22}~
keV\cdot b \hfill\cr
S_{33}(0)\hfill &
4.99\times {10}^{3}~keV\cdot b \hfill &5.6\times {10}^{3}~keV\cdot b 
\hfill \cr S_{34}(0)\hfill &0.524~keV\cdot b \hfill & 
0.45~keV\cdot b\hfill \cr S_{17}(0)\hfill&0.0224~keV\cdot b \hfill
& 0.017~keV\cdot b\hfill \cr
{\rm Screening~ Effects}\hfill &{\rm Included } \hfill&{\rm Included} 
\hfill \cr 
{\rm Nuclear~Equilibrium}\hfill&{\rm Imposed} \hfill&{\rm Not~Assumed} 
\hfill \cr}$$
\clearpage

\clearpage
 
{\bf Table III:} Fractional change in the predicted $\nu_\odot$ fluxes 
and counting rates in the $\nu_\odot$ experiments due to the inclusion of 
element diffusion in 
the SSM calculations of Bahcall and Pinsonneault (1996), Dar and Shaviv 
(1994, 1996) and Richard, Vauclair, Charbonnel and Dziembowski (1996).
The results of models 1 and 2 of RVCD were   
extrapolated to the initial solar composition which was used in DS96. 
 
$$\matrix{\phi_{\nu_\odot} & {\rm BP95}& {\rm DS96} & {\rm RVCD}\cr
pp\hfill &-~1.7\%\hfill &-~0.3\%\hfill &-~0.8\% \hfill\cr
pep\hfill &-~2.8\%\hfill &-~0.3\%\hfill &-~0.4\%\hfill\cr
{\rm ^7Be}\hfill    &+13.7\%\hfill &+4.2\%\hfill  &+ ~6.5\%\hfill\cr
{\rm ^8B}\hfill    &+36.5\%\hfill &+11.2\%\hfill &+10.7\%\hfill\cr
{\rm ^{13}N}\hfill &+51.8\%\hfill &+22.7\%\hfill &+19.8\%\hfill\cr
{\rm ^{15}O}\hfill &+58.0\%\hfill &+24.0\%\hfill &+20.8\%\hfill\cr
{\rm ^{17}F}\hfill &+61.2\%\hfill &+24.9\%\hfill &+21.8\%\hfill\cr
{\rm Rates} & & & {\rm RVCD}\cr
{\rm H2O}\hfill &+36.5\%\hfill &+11.2\%\hfill&+13.3\%\hfill\cr
{\rm Cl}\hfill &+32.9\%\hfill &+~9.5\%\hfill&+12.3\%\hfill\cr
{\rm Ga}\hfill &+~8.7\%\hfill &+~2.6\%\hfill&+~3.7\%\hfill
\cr}$$
 
\vfill
\eject
 
\end{document}